\documentclass[preprint,5p,number,a4paper,sort&compress]{elsarticle}

\usepackage{graphicx}
\usepackage{epstopdf}
\usepackage{dcolumn}
\usepackage{amsmath}
\usepackage{mathptmx, courier, pifont}
\usepackage[scaled=0.92]{helvet}
\usepackage[T1]{fontenc}
\usepackage{textcomp}
\usepackage{color}
\usepackage{lineno,hyperref}
\modulolinenumbers[1]

\journal{Nuclear Instruments and Methods in Physics Research Section A}

\begin{document}
\begin{frontmatter}

\title{A method to measure the transition energy $\gamma_{t}$ of the isochronously tuned storage ring}

\cortext[mycorrespondingauthor]{Corresponding authors}
\author[A]{R.~J.~Chen}
\author[A]{X.~L.~Yan}
\author[A,B]{W.~W.~Ge}
\author[A]{Y.~J.~Yuan\corref{mycorrespondingauthor}}
\ead{yuanyj@impcas.ac.cn}
\author[A,F]{M.~Wang\corref{mycorrespondingauthor}}
\ead{wangm@impcas.ac.cn}
\author[A]{M.~Z.~Sun}
\author[A]{Y.~M.~Xing}
\author[A,B]{P.~Zhang}
\author[A,B]{C.~Y.~Fu}
\author[A]{P.~Shuai}
\author[A]{X.~Xu}
\author[A,F]{Y.~H.~Zhang}
\author[A]{T.~Bao}
\author[A,C]{X.~C.~Chen}
\author[A]{X.~J.~Hu}
\author[A,e]{W.~J.~Huang}
\author[A,B]{H.~F.~Li}
\author[A,B]{J.~H.~Liu}
\author[A,C]{Yu.~A.~Litvinov}
\author[A,C]{S.~A.~Litvinov}
\author[A]{L.~J.~Mao}
\author[A]{B.~Wu}
\author[A,F]{H.~S.~Xu}
\author[A]{J.~C.~Yang}
\author[A]{D.~Y.~Yin}
\author[A,D]{Q.~Zeng}
\author[A]{X.~H.~Zhang}
\author[A]{W.~H.~Zheng}
\author[A,F]{X.~H.~Zhou}
\author[A]{X.~Zhou}

\address[A] {CAS Key Laboratory of High Precision Nuclear Spectroscopy and \\Center for Nuclear Matter Science, Institute of Modern Physics, Chinese Academy of Sciences, Lanzhou 730000, China}
\address[B] {University of Chinese Academy of Sciences, Beijing, 100049, People's Republic of China}
\address[C] {GSI Helmholtzzentrum f{\"u}r Schwerionenforschung, Planckstra{\ss }e 1, 64291 Darmstadt, Germany}
\address[D] {Research Center for Hadron Physics, National Laboratory of Heavy Ion Accelerator Facility in Lanzhou and \\ University of Science and Technology of China, Hefei 230026, China}
\address[e] {CSNSM, Univ Paris-Sud, CNRS/IN2P3, Universit\'{e} Paris-Saclay, 91405 Orsay, France.}
\address[F] {Joint Research Center for Modern Physics and Clean Energy, South China Normal University, Institute of Modern Physics, Chinese Academy of Sciences, Lanzhou 730000, People's Republic of China.}

\begin{abstract}
The Isochronous Mass Spectrometry (IMS) is a powerful technique developed in heavy-ion storage rings for measuring masses of very short-lived exotic nuclei.
The IMS is based on the isochronous setting of the ring. One of the main parameters of this setting is the transition energy $\gamma_{t}$.
It has been a challenge to determine the $\gamma_{t}$ and especially to monitor the variation of $\gamma_{t}$ during experiments.
In this paper we introduce a method to measure the $\gamma_{t}$ online during IMS experiments by using the acquired experimental data.
Furthermore, since the storage ring has (in our context) a relatively large momentum acceptance,
the variation of the $\gamma_{t}$ across the ring acceptance is a source of systematic uncertainty of measured masses.
With the installation of two time-of-flight (TOF) detectors, the velocity of each stored ion and its revolution time are simultaneously available for the analysis.
These quantities enabled us to determine the $\gamma_{t}$ as a function of orbital length in the ring.
The presented method is especially important for future IMS experiments
planned at the new-generation storage ring facilities FAIR in Germany and HIAF in China.
\end{abstract}


\end{frontmatter}
\section{Introduction}
Masses are basic nuclear properties.
Their accurate knowledge is particularly important for nuclear structure and nuclear astrophysics \cite{Lunney2003,Blaum2006}.
The challenge today is to obtain accurate masses of nuclei located far away from the valley of $\beta$-stability \cite{100yms}.
However, such nuclei are as a rule short-lived and are produced with tiny yields.
Therefore, highly efficient and fast measurement techniques are required.
The Isochronous Mass Spectrometry (IMS) is one of such techniques
which is realised at in-flight radioactive ion beam facilities \cite{Litvinov2010,Bosch2013,ZhangPS2016}.
The IMS experiments are performed today at three heavy-ion storage ring facilities,
namely at the experimental storage ring ESR at GSI Helmholtz Center in Darmstadt
\cite{Hausmann2001, Stadlmann2004, Sun2008, KnobelEPJ2016},
at the experimental cooler-storage ring CSRe at the Institute of Modern Physics in Lanzhou
\cite{Meng2009,TuNIMA2011,TuPL2011,ZhangPRL2012,YanAPL2013,ShuaiPLB2014,Xu2016,Zhang2017},
and at the rare-ion storage ring R3 at the RIKEN Nishina Center in Tokyo
\cite{Ozawa2012,Yamaguchi2013a,Yamaguchi2013b,Wakasugi2015,Yamaguchi2015a,Yamaguchi2015b}.
The discussion in this paper is based on the experimental results obtained at the CSRe \cite{XiaNIMA2002}.
However, the conclusions are valid for IMS experiments at all present and future storage ring facilities \cite{ZhangPS2016}.

The IMS is based on a special --isochronous-- tuning of the storage ring
\cite{WollnikNIMA1987,HausmannNIMA2000,Franzke2008}.
The principle of the IMS is expressed by the following equation (\ref{eqims}),
which connects the revolution times ($T$) of stored ions in a storage ring and their variations ($\Delta T$)
to the differences in mass-to-charge ratios ($m/q$) and differences in velocity ($v$)
\cite{WollnikNIMA1987,HausmannNIMA2000,Franzke2008}:

\begin{equation}
   \frac{\Delta T}{T} = \frac{1}{{\gamma_{t}}^{2}}\frac{\Delta (m/q)}{m/q} - \Biggl(1-\frac{{\gamma}^{2}}{{\gamma}^{2}_{t}}\Biggr)\frac{\Delta v}{v},
   \label{eqims}
\end{equation}
where $\gamma$ is the relativistic Lorentz factor and $\gamma_{t}$ is the transition energy of the storage ring.
The tuning of the storage ring into the isochronous mode means
that the revolution times of the stored ions are independent from their velocity.
The necessary condition for the latter is that the energy of the ions corresponds to $\gamma = \gamma_{t}$.
In this mode, the revolution times of ions with same $m/q$ but with different velocities $v$
are compensated by different lengths of their orbits \cite{HausmannNIMA2000}.
If the second term in equation (\ref{eqims}) can be neglected
then the revolution time ($T$) depends only on the mass-to-charge ratio ($m/q$).
This is the basis of the IMS mass measurements \cite{Franzke2008}.

Due to the fixed magnetic rigidity, $B\rho=mv\gamma/q$, of the transport lines and the injection into the ring,
nuclides with different $m/q$ values inevitably have different mean velocities.
In a realistic experiment, the simultaneously covered $\Delta{(m/q)}/(m/q)$ range is about 10\%,
and thus the treatment of ions with $\gamma \ne \gamma_{t}$ is an essential issue.
Furthermore, the emittance of the secondary ions after the nuclear production reaction is --as a rule-- larger than the acceptance of the storage ring \cite{Geissel1995}.
Thus, the knowledge of the $\gamma_{t}$ as a function of storage ring acceptance is also important.

If the isochronous condition $\gamma = \gamma_{t}$ is not strictly fulfilled within the momentum acceptance of the storage ring,
the revolution time spread ${\Delta T}/{T}$ is determined
by the momentum acceptance of the storage ring ${\Delta P}/{P}$ and the slip factor, $\eta$.
For the case of the CSRe, the ${\Delta P}/{P} \sim 0.1\%$ \cite{XiaNIMA2002}.
For the ions with same $m/q$, equation (\ref{eqims}) can be rewritten as:
\begin{equation}
   \frac{\Delta T}{T} = - \eta\frac{\Delta P}{P}= - \Biggl(\frac{1}{{\gamma}^{2}}-\frac{1}{{\gamma}^{2}_{t}}\Biggr)\frac{\Delta P}{P}.
   \label{eqims2}
\end{equation}

The mass resolving power is directly connected to ${\Delta T}/{T}$.
There are various methods to reduce ${\Delta T}/{T}$ \cite{ChenRST2015}.
One method is to tune the $\gamma_{t}$ in dependence to ${\Delta P}/{P}$
such that the $\eta$ parameter is minimal \cite{DolinskiiNIMA2007,DolinskiiNIMA2008,LitvinovNIMA2013}.
This method is especially important for ions with $\gamma \approx \gamma_{t}$.

Another method is to reduce the ring injection acceptance ${\Delta P}/{P}$ by using a mechanical slit.
This is the so-called $B\rho$-tagging method \cite{KnobelEPJ2016,Geissel2006,KnobelPLB2016}.
Indeed, the mass resolving power can significantly be increased  at a cost of a dramatically reduced transmission.
The later hinders any practical application of the method in mass measurements of rarely produced nuclides.
The experiments employing the $B\rho$-tagging have shown that additional measurement of velocities or magnetic rigidities of each ion
can be used to reduce ${\Delta T}/{T}$ also for non-isochronous particles \cite{Geissel2005}.

In all the approaches above the knowledge of $\gamma_{t}$ as a function of the ring acceptance is indispensable.
However, up to now it has been a challenge to deduce the $\gamma_{t}$.

An effective method developed at the ESR uses the electron cooling of primary beam ions \cite{HausmannNIMA2000}.
The velocity of the particles is defined by the cooler to better than $\Delta{v}/v\approx10^{-5}$.
The storage ring acceptance is scanned by changing the ion velocity
and the corresponding revolution frequencies are used to extract the $\gamma_{t}$.
Since the $m/q$-values of the ions of interest are not equal to the $m/q$ of the primary beam ions,
the setting of the storage ring is not isochronous for the primary beam ions.
The extraction of the $\gamma_{t}$ is possible assuming that the particles
with the same magnetic rigidity travel on exactly the same orbits in the ring \cite{HausmannNIMA2000,YanPST2015}.
It is also clear that this method can not be used for online monitoring of $\gamma_{t}$.

At the CSRe, the information on the $\gamma_{t}$ is obtained from the widths of peaks in the measured revolution time spectra.
However, this method is largely affected by the instabilities of magnetic fields of ring magnets.
Although, various corrections of the magnetic field instabilities can be applied \cite{TuNIMA2011},
only the $\gamma_{t}$-value averaged over the ring acceptance could so far be accessed.

In this paper, we introduce a new method which can be used to determine the $\gamma_{t}$ precisely by using measured data.
The measurements of velocities of each stored ion has been recently implemented at the CSRe
\cite{XingPST2015,XuCPC2015,XuNIMB2016}.
On the one hand, this allowed us improving the mass resolving power of the IMS without losing valuable statistics.
On the other hand, it is possible to measure the $\gamma_{t}$ as a function of storage ring acceptance.

\section{Basic equations}

In the IMS experiments, the revolution times are measured by dedicated time-of-flight (ToF) detectors
\cite{Troetscher1992,MeiNIMA2010}.
Such detector is equipped with a thin foil which is penetrated by revolving ions at each revolution.
In the CSRe ToF detector a 20 $\mu g/cm^{2}$ carbon foil is used \cite{MeiNIMA2010,ZhangNIMA2014}.
Secondary electrons released from the foil surface due to the passing  ions are guided isochronously
by perpendicularly arranged electrostatic and magnetic fields to a set of micro-channel plates (MCP).
The signals from the MCPs \cite{MeiNIMA2010,ZhangNIMA2014} are directly input into a fast oscilloscope.
In the discussed experiments we employed a digital oscilloscope at a sampling rate of 50 GHz.
The detection efficiency of each ion at every revolution varied from 20 to 70$\%$,
depending on the charge of the ions and the overall number of stored ions.
The recording time was 200 $\mu$s triggered by the injection of the ions into the CSRe,
which corresponds to $\sim$300 revolutions.
Several tens of signals could be obtained for each ion during the recording time.
They were used to determine the mean revolution time.
For this purpose the signals from each ion were fitted by a second order polynomial function
with fit coefficients $A_0$, $A_1$, and $A_2$.
\cite{TuPL2011,ZhangPRL2012,YanAPL2013,ShuaiPLB2014,TuPST2015}:

\begin{equation}
   time = A_{0} + A_{1}Turn + A_{2}Turn^{2},
   \label{timeturn}
\end{equation}
where $time$ stands for the time after the injection of ions into the ring as measured by the oscilloscope
and $turn$ indicates the revolution number.
The revolution time of the ion is determined as a slope of the fit function (\ref{timeturn}) at a given turn number:

\begin{equation}
   T(Turn) = \frac{d(time)}{d(Turn)} = A_{1} + 2A_{2}Turn
   \label{T_Turn}
\end{equation}
It is clear from equation (\ref{T_Turn}) that the extracted revolution time is different at different turn numbers.
The main reason for this is the energy loss of particles in the carbon foil of the TOF detector.
The difference of revolution times in two adjacent turns can be written as:

\begin{equation}
   \delta T = T(Turn+1) - T(Turn) = 2A_{2}.
   \label{2A2}
\end{equation}

The relative momentum change of an ion ${\delta P}/{P}$
due to the passage through the detector foil is connected to the change in kinetic energy, ${\delta E_{k}}/{E_{k}}$, as:

\begin{equation}
   \frac{\delta P}{P} = \frac{\gamma}{\gamma + 1} \frac{\delta E_{k}}{E_{k}},
   \label{dpop_dEkEk}
\end{equation}
where $E_{k}$ and $\delta E_{k}$ are the kinetic energy and an average energy loss in the foil.
The energy loss of ions at each revolution follows the Landau distributions.
Since the measurements are performed for several hundreds revolutions, the $\delta E_{k}$ represents the average energy loss.

By combining equations (\ref{dpop_dEkEk}) and (\ref{eqims2}) we can rewrite:

\begin{equation}
   \frac{\delta T}{T} = - \Biggl(\frac{1}{{\gamma}^{2}}-\frac{1}{{\gamma}^{2}_{t}}\Biggr)\frac{\gamma}{\gamma + 1} \frac{\delta E_{k}}{E_{k}}.
   \label{dToT_dEkEk}
\end{equation}

By inserting equation (\ref{2A2}) into equation (\ref{dToT_dEkEk}), we can get the expression for $A_2$ coefficient:

\begin{equation}
   A_{2} = - \frac{1}{2}\bigg(\frac{1}{{\gamma}^{2}}-\frac{1}{{\gamma}^{2}_{t}}\bigg)\frac{\gamma}{\gamma + 1} \frac{\delta E_{k}}{E_{k}}T.
   \label{A2Formular}
\end{equation}

According to equation (\ref{A2Formular}),  the $A_{2}$ coefficient is determined by the parameters of the ion
(the relativistic Lorentz factor $\gamma$, the average energy loss $\delta E_{k}$, the kinetic energy $E_{k}$, and the revolution time $T$)
and by the ring parameter $\gamma_{t}$.

In the case of the CSRe set into the isochronous mode with $\gamma_{t}$=1.395,
a reference --isochronous-- ${}^{52}$Co$^{27+}$ ion with $m/q$ = 1.9240126(3) \cite{Huang2016}
has the kinetic energy $E_k=372$~A~MeV.
The average energy loss due to a 20~$\mu$g/cm$^{2}$ carbon foil is about $\delta{E_{k}}=0.0416$~MeV (0.0008~A~MeV) \cite{LISE}.
With the revolution time $T = 614184.9$~ps, corresponding to the central orbit length of 128801~mm \cite{XiaNIMA2002},
$A_{2}$ can be estimated with equation (\ref{A2Formular}) to be about 0.0012~ps/$Turn^{2}$.

\section{Comparison with experimental data}
\begin{figure}\centering
	\includegraphics[angle=0,width=8.5 cm]{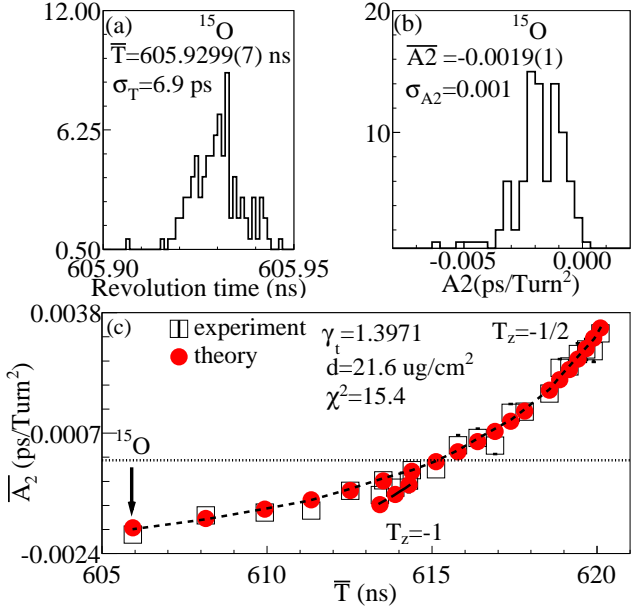}
	\caption{(Colour online)
	Panels (a) and (b) show the distributions of revolution times and $A_{2}$ coefficients for $^{15}$O ions, respectively.
	Experimental and calculated with equation (\ref{A2Formular}) average $A_{2}$ coefficients as a function
	of the revolution time of stored ions are shown in panel (c).
	The black open rectangles indicate the experimental data acquired for one hour.
	The red filled circles indicate the calculated results for $\gamma_{t}$=1.3971 and the thickness of carbon foil $d = 21.6$~$\mu$g/cm$^{2}$.
	The upper and lower dashed lines are for nuclei with $T_{z}=(N-Z)/2$=$-$1/2 and $T_{z}$=$-$1, respectively.
	The isochronous setting of the CSRe was optimised for ions around 615 ns.
\label{Ni58ExpOneHour}}
\end{figure}

Each stored nuclide has its specific revolution time determined by its mass-over-charge ratio.
The distribution of revolution times and A$_{2}$ for $^{15}$O ions are shown in Figure \ref{Ni58ExpOneHour} (a) and (b), respectively.
The  revolution times spread for $^{15}$O ions is about $\sigma_{T}$=6.9 ps and the average revolution time $\overline{T}$ is 605.9299(7) ns.
The  A$_{2}$ spread for $^{15}$O ions is about $\sigma_{A2}$=0.001 ps/Turn$^{2}$ and the average A$_{2}$  is $-$0.0019(1) ps/Turn$^{2}$.
The relation between average revolution time of each nuclide and its corresponding average $A_{2}$ coefficient is shown in Figure \ref{Ni58ExpOneHour} (c).
The open rectangles represent the experimental data obtained for one hour at the CSRe \cite{Xu2016}.
The red filled circles represent results calculated by using equation (\ref{A2Formular})
assuming $\gamma_{t}$=1.3971 and the thickness of carbon foil $d$ = 21.6 $\mu g/cm^{2}$.
It is noted that the $\gamma$ of each ion species was calculated by its average revolution time $\overline{T}$ and central orbit $C=128801$ mm.
The arrow indicates the $^{15}$O ions.
Nuclei belonging to different isospin series, characterised by $T_{z}=(N-Z)/2$, have different transmissions and mean kinetic energies \cite{TuPRC2016}.
Thus, the upper and lower dashed lines are for nuclei with $T_{z}$=$-$1/2 and and $T_{z}$=$-$1, respectively.
The absolute value of the average $\overline{A_{2}}$ is close to 0 around $\overline{T} = 615$ ns and becomes larger when moving away from it.
The calculations qualitatively agree with the experimental data.
To quantify the agreement, the $\chi^{2}$ can be used. It is defined as:

\begin{equation}
   \chi ^{2} = \sum_{i=1}^{N} \frac{(\overline{A_{2,i}^{exp.}} - A_{2,i}^{theo.})^{2}}{e_{\overline{A_{2,i}^{exp.}}}^{2}}.
   \label{Chi2}
\end{equation}
Here $i$ presents the $i^{th}$ nuclide and $N$ presents the total number of nuclei in the comparison.
The $\overline{A_{2,i}^{exp.}}$ and $A_{2,i}^{theo.}$ are the experimental average and theoretical  $A_{2}$ coefficient for the $i^{th}$ nuclide.
The $e_{\overline{A_{2,i}^{exp.}}}$ value is the uncertainty of  $\overline{A_{2,i}^{exp.}}$ value for the $i^{th}$ nuclide.
The $\chi ^{2}$=15.4 was obtained for the data illustrated in Figure~\ref{Ni58ExpOneHour}.

\begin{figure}\centering
	\includegraphics[angle=0,width=8.5 cm]{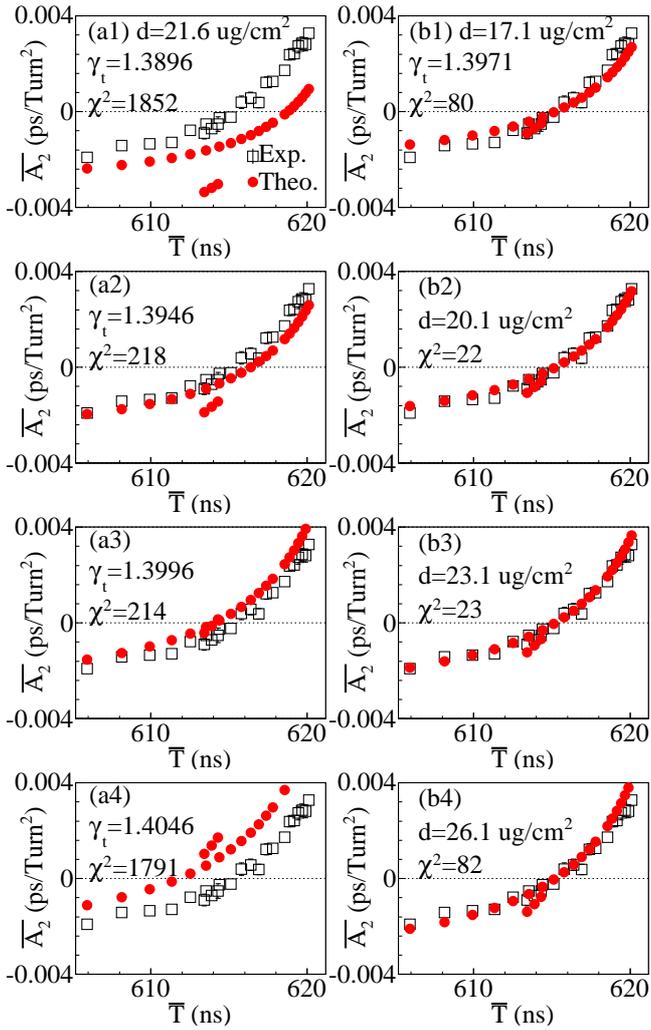}
	\caption{(Colour online)
	The same as Figure~\ref{Ni58ExpOneHour}, but (i) for
	the fixed foil thickness $d=21.6$ $\mu$g/cm$^2$ and four different $\gamma_{t}=1.3896,$ 1.3946, 1.3996, and 1.4046 (panels (a1)-(a4), respectively),
	and (ii) for the fixed $\gamma_{t}=1.3971$ and four different foil thicknesses $d=17.1$, 20.1, 23.1, and 26.1 $\mu$g/cm$^2$ (panels (b1)-(b4), respectively).
	\label{Ni58ExpOneHour_DifferentCondition}}
\end{figure}

The sensitivity of calculations to the thickness of carbon foil $d$ and the $\gamma_{t}$ can be tested.
Four left panels of Figure~\ref{Ni58ExpOneHour_DifferentCondition} (a1)-(a4) show the calculations performed
for four $\gamma_{t}$ values varied from 1.3896 to 1.4046 in steps of 0.0005 and a fixed thickness of carbon foil $d=21.6$~$\mu$g/cm$^{2}$.
Since the kinetic energy of the ions is high, the average energy loss can safely be assumed to be proportional to the thickness of the foil.
Four right panels of Figure~\ref{Ni58ExpOneHour_DifferentCondition} (b1)-(b4) show the calculations performed
for four thicknesses of carbon foil varied from $d=17.1$~$\mu$g/cm$^{2}$ to $d=26.1$~$\mu$g/cm$^{2}$ in steps of 3~$\mu$g/cm$^{2}$
and a fixed $\gamma_{t}=1.3971$.

\begin{figure}\centering
	\includegraphics[angle=0,width=8.5 cm]{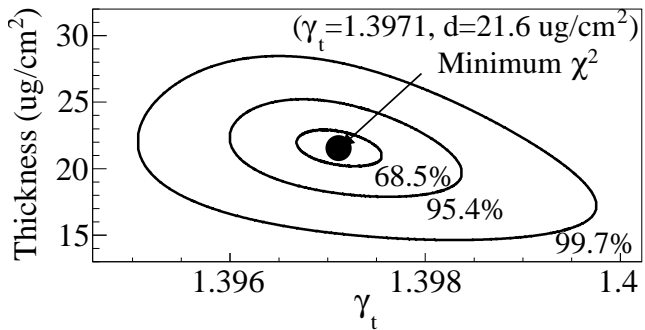}
	\caption{
	Agreement between experimental and calculated data assuming various thicknesses of carbon foil and the $\gamma_{t}$ values.
	The $\chi^{2}$ is defined by equation~(\ref{Chi2}).
	The three ellipses represent contours for confident levels of 68.5\%, 95.4\% and 99.7\%.
	The black point indicates the overall $\chi^{2}$ minimum.
	\label{Contours}}
\end{figure}

It can be seen that the description of the experimental data is very sensitive to the assumed thickness of the foil $d$ and $\gamma_{t}$.
One can employ this sensitivity to estimate both parameters from the measured data.
Figure~\ref{Contours} shows the $\chi^{2}$ values for different $d$ and $\gamma_{t}$ as a contour plot.
The three shown contours illustrate three confidence levels of 68.5$\%$, 95.4$\%$ and 99.7$\%$.
From the minimum of the $\chi^{2}$ distribution and one standard deviation ellipse, the thickness of carbon foil
$d$ = 21.6$\pm$1.4~$\mu$g/cm$^{2}$ and $\gamma_{t}$=1.3971$\pm$0.0005 can be deduced.

\section{Variation of $\gamma_{t}$ in time}
Due to the instabilities of magnetic fields of the storage ring magnets, the $\gamma_{t}$ can vary in time.
Therefore, the measured $\gamma_{t}$ represents the $\gamma_{t}$ averaged over the measurement period.
Furthermore, there is no information on the dependence of the $\gamma_{t}$ as a function of the storage ring acceptance.

\begin{figure}\centering
	\includegraphics[angle=0,width=8.5 cm]{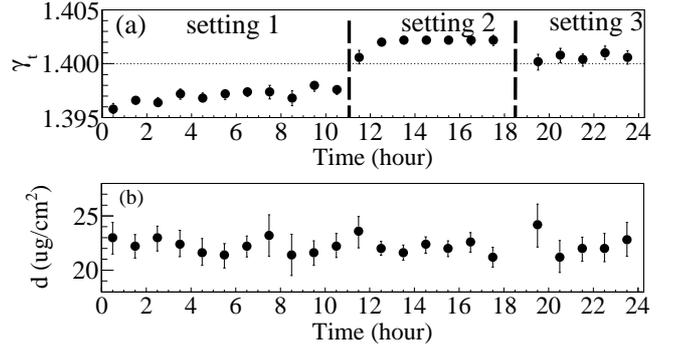}
	\caption{(Colour online)  The dependence of $\gamma_{t}$ (a) and the thickness of carbon foil $d$ (b) as a function of time.
	\label{gammat_time}}
\end{figure}

Figure~\ref{gammat_time} shows the behavior of $d$ and $\gamma_{t}$ over one day measurement time in one-hour steps
as measured for in a recent experiment where the ${}^{52}$Co$^{27+}$ ions was set as the reference for the isochronous setting \cite{Xu2016}.
The $d$ and $\gamma_{t}$ parameters are deduced at each step from the corresponding minimal $\chi^{2}$ values.
During the time period illustrated in Figure~\ref{gammat_time}, two adjustments of the CSRe were performed.
At first, the setting of CSRe quadrupoles was modified at about 11:00 time mark,
which is indicated as ``Setting~1'' in Figure~\ref{gammat_time}.
This modification caused the change of $\gamma_{t}$ from 1.396 to 1.402.
The latter is indicated as ``Setting~2'' in Figure~\ref{gammat_time}.
At about 18:30 time mark, a mechanical slit in the CSRe has been implemented, ``Setting-3'',
which reduced the storage acceptance of the ring.
This caused the $\gamma_{t}$ change from 1.402 to 1.400.
The latter change indicates that the $\gamma_{t}$ is not constant over the acceptance of the CSRe.
Although the average foil thickness shall not change in time and the value of $d$ can be fixed in the calculations,
it will be shown that the developed tool enables us for testing the homogeneity of the foil thickness over the foil area.

\begin{figure}\centering
	\includegraphics[angle=0,width=8.5 cm]{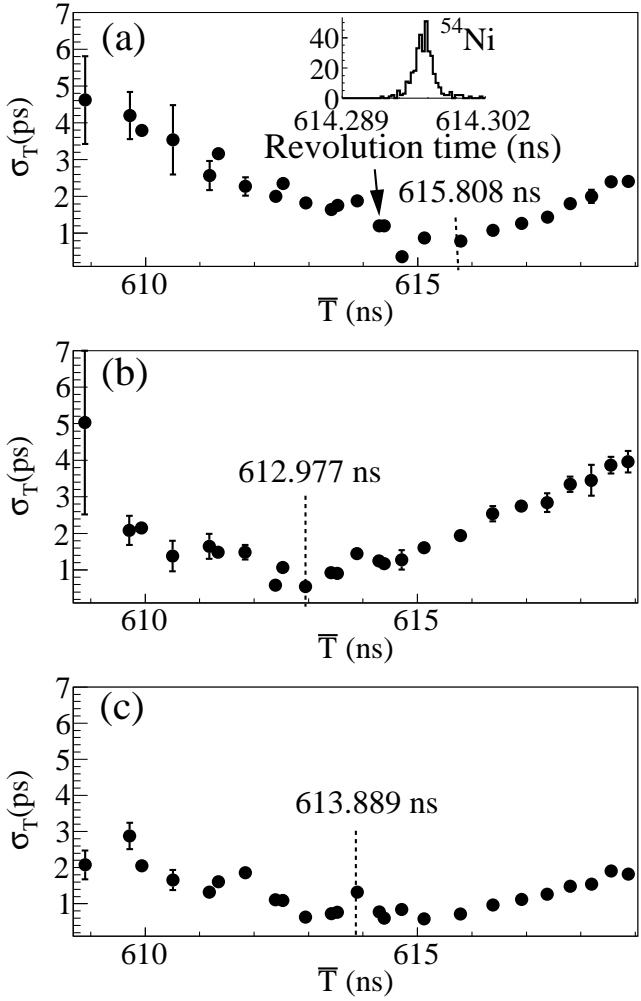}
	\caption{
	Standard deviations, $\sigma_t$, of revolution time distributions of different nuclides as a function of their mean revolution times.
	Panels (a), (b) and (c) show the data for settings 1, 2 and 3 (see Figure \ref{gammat_time}), respectively.
	The insert in (a) shows the revolution time distribution for ${}^{54}$Ni ions.
	The dashed lines indicate the corresponding isochronous settings.
	\label{gammat_T}}
\end{figure}

The characterisation of isochronous settings is conventionally done
by plotting the standard deviations, $\sigma_t$, of revolution time distributions of all nuclides versus their mean revolution times.
This is illustrated in Figure \ref{gammat_T} where the data for settings 1-3 are presented.
The best isochronous ions are found at 615.808~ns, 612.977~ns and 613.889~ns, respectively.
By taking into account the length of the CSRe central orbital (128801~mm),
the corresponding $\gamma_{t}$ values are 1.396, 1.402, and 1.400.
These rough estimated $\gamma_{t}$ values agree well with the ones shown in Figure \ref{gammat_time}.
Although the $\gamma_{t}$ changes for different settings, the foil thickness stays within uncertainties, as expected, almost constant.
The calculated thickness of the foil is 22.3$\pm$1.3~$\mu$g/cm$^{2}$ .

\section{Variation of $\gamma_{t}$ over the ring acceptance}

With the installation of two ToF detectors in the CSRe, the measurements of velocities of each stored ions are enabled.
This allows us to independently evaluate the dependence of $\gamma_{t}$ on the $\Delta{P}/P$ or
equivalently on the orbital pass length, $C$.

\begin{figure}\centering
	\includegraphics[angle=0,width=8.5 cm]{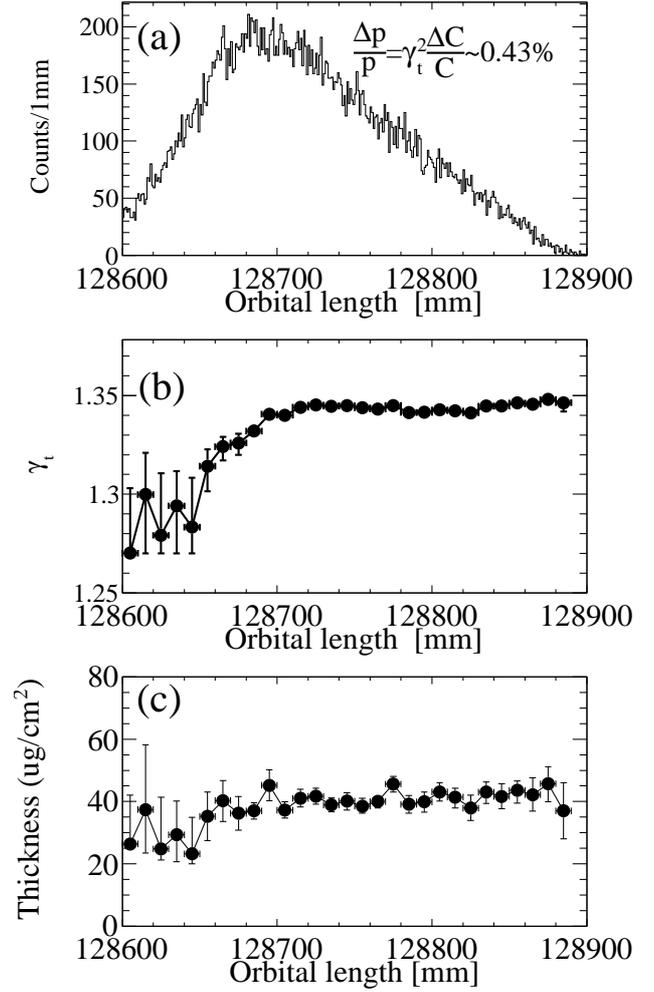}
	\caption{
	The number of injected ions (a), the $\gamma_{t}$ (b) and the foil thickness $d$ (c) as a function of orbital length $C^\prime$ (see text).
	\label{DistributionOfC}}
\end{figure}

The accuracy of the time determination of a single ToF detector is $\sim$18~ps.
The distance between the two ToF detectors is 18034(1)~mm, corresponding to the time-of-flight of about 87~ns.
The relative accuracy of the velocity measurement is about $1 \times 10^{-4}$.
The orbital length for each ion can be calculated with \cite{XuCPC2015}:

\begin{equation}
   C = v T = \frac{L}{t_{2}-t_{1}-\Delta t_{cable2-1}} T,
   \label{CFormula}
\end{equation}
where $C$, $v$, $T$, $L$, $t_{2}$, $t_{1}$,$\Delta t_{cable2-1}$ are
the orbital length, velocity, revolution time, distance between the two ToF detectors,
time measured by the second ToF detector, time measured by the first ToF detector,
and time delay due to the different cable lengths between two TOF detectors and the oscilloscope, respectively.

The exact value of $\Delta t_{cable2-1}$ is not available for the presented measurements.
It will be accurately determined in the future.
In the calculation of the orbital length here we used $\Delta t_{cable2-1} = 0$ ps,
which causes the calculated orbit lengths, $C^\prime$, to be shifted from their real values, $C$.
However, the shape of the $\gamma_{t}$ as a function of orbital length will be preserved.

The distribution of injected ions as a function of the orbital length is shown in Figure \ref{DistributionOfC} (a).
All available data are subdivided into groups according to their orbital lengths.
Each bin is about 10~mm.
The average $\gamma_{t}$ and the foil thickness $d$ can be determined for each of these groups.
It can be seen in Figure~\ref{DistributionOfC} (b)
that the $\gamma_{t}$ stays nearly constant in the range of orbit lengths from 128700~mm to 128900~mm, and
that it changes rapidly for lower orbit lengths.

The thickness of the carbon foil as a function of the orbital length is shown in Figure~\ref{DistributionOfC} (c).
The mean value is estimated to be 39.2$\pm$6~$\mu$g/cm$^{2}$,
which is reasonable due to the two ToF detectors with two foils of $\sim$20~$\mu$g/cm$^{2}$ each.
This technique may be helpful for measurements of the thicknesses of ultra-thin solid foils or internal targets or even rest-gas pressure.
Since the orbital length $C$ can be translated into the transverse coordinate,
the variation of the material thickness in its radial direction might be possible to deduce.

\section{Conclusion}
An analysis of the acquired experimental data from IMS mass measurements has been performed.
The second coefficient, $A_{2}$, of the parabolic fit of the measured time versus the revolution number of the ion in the ring
is determined by the relativistic Lorentz factor $\gamma$, the kinetic energy $E_{k}$,
revolution time $T$, and the average energy loss $\Delta E_{k}$ of the ion as well as by the transition energy of the ring $\gamma_{t}$.
By comparing the calculated and experimental $A_{2}$ coefficients,
the  average $\gamma_{t}$ and the thickness of the detector foil $d$ can accurately be obtained.
By combining the velocity measurement of each stored ion,
which became recently enabled in the IMS experiments with two time-of-flight detectors,
the dependence of the $\gamma_{t}$ as a function of orbital length $C$ can be obtained.
Since the online data can directly be used for such investigations,
the developed method is expected to be used as an online monitor of the $\gamma_{t}$.
In addition, this method may be applied to measure thicknesses of ultra-thin materials.

The presented method is of importance for the future experiments at the presently running storage rings facilities \cite{Litvinov2013}.
Furthermore, it is indispensable for the IMS experiments planned
at the next-generation radioactive-ion beam facilities FAIR in Germany \cite{FAIR} and HIAF in China \cite{HIAF}.
For instance, the IMS measurements are planned in the collector ring, CR, at FAIR \cite{ILIMA}, where no electron cooling is foreseen.
The determination of the $\gamma_{t}$ was proposed based on the variation of the primary beam energy in the main accelerator.
Since the installation of two ToF detectors in the CR is foreseen, the characterisation of the $\gamma_{t}$ will be straightforward by using the developed here method.

\section*{Acknowledgments}
This work is supported in part by National Key R$\&$D Program of China (Contract No.  2016YFA0400504),
the NSFC (Grant Nos. 11605252, U1232208, U1432125, 11205205, 11035007,  11605248),
the Helmholtz-CAS Joint Research Group HCJRG-108,
the External Cooperation Program of the CAS (GJHZ1305),
the 973 Program of China (No. 2013CB834401),
and by the European Research Council (ERC) under the European Union's Horizon 2020 research and innovation programme (grant agreement No 682841 "ASTRUm").

\end{document}